\documentclass[twocolumn, switch]{article}

\usepackage{arxiv}

\usepackage[utf8]{inputenc} 
\usepackage[T1]{fontenc}    
\usepackage{url}            
\usepackage{booktabs}       
\usepackage{amsfonts}       
\usepackage{amsmath}        
\usepackage{nicefrac}       
\usepackage{microtype}      
\usepackage{lineno}         
\usepackage{setspace}       
\usepackage{graphicx}       
\usepackage{algorithm}
\usepackage{multirow}

\usepackage[draft]{minted}
\usepackage{subcaption}
\usepackage[font=small]{caption}
\graphicspath{{figures/}}
\usepackage{doi}

\makeatletter
\def\PYG@reset{\let\PYG@it=\relax \let\PYG@bf=\relax%
    \let\PYG@ul=\relax \let\PYG@tc=\relax%
    \let\PYG@bc=\relax \let\PYG@ff=\relax}
\def\PYG@tok#1{\csname PYG@tok@#1\endcsname}
\def\PYG@toks#1+{\ifx\relax#1\empty\else%
    \PYG@tok{#1}\expandafter\PYG@toks\fi}
\def\PYG@do#1{\PYG@bc{\PYG@tc{\PYG@ul{%
    \PYG@it{\PYG@bf{\PYG@ff{#1}}}}}}}
\def\PYG#1#2{\PYG@reset\PYG@toks#1+\relax+\PYG@do{#2}}

\@namedef{PYG@tok@w}{\def\PYG@tc##1{\textcolor[rgb]{0.73,0.73,0.73}{##1}}}
\@namedef{PYG@tok@c}{\let\PYG@it=\textit\def\PYG@tc##1{\textcolor[rgb]{0.25,0.50,0.50}{##1}}}
\@namedef{PYG@tok@cp}{\def\PYG@tc##1{\textcolor[rgb]{0.74,0.48,0.00}{##1}}}
\@namedef{PYG@tok@k}{\let\PYG@bf=\textbf\def\PYG@tc##1{\textcolor[rgb]{0.00,0.50,0.00}{##1}}}
\@namedef{PYG@tok@kp}{\def\PYG@tc##1{\textcolor[rgb]{0.00,0.50,0.00}{##1}}}
\@namedef{PYG@tok@kt}{\def\PYG@tc##1{\textcolor[rgb]{0.69,0.00,0.25}{##1}}}
\@namedef{PYG@tok@o}{\def\PYG@tc##1{\textcolor[rgb]{0.40,0.40,0.40}{##1}}}
\@namedef{PYG@tok@ow}{\let\PYG@bf=\textbf\def\PYG@tc##1{\textcolor[rgb]{0.67,0.13,1.00}{##1}}}
\@namedef{PYG@tok@nb}{\def\PYG@tc##1{\textcolor[rgb]{0.00,0.50,0.00}{##1}}}
\@namedef{PYG@tok@nf}{\def\PYG@tc##1{\textcolor[rgb]{0.00,0.00,1.00}{##1}}}
\@namedef{PYG@tok@nc}{\let\PYG@bf=\textbf\def\PYG@tc##1{\textcolor[rgb]{0.00,0.00,1.00}{##1}}}
\@namedef{PYG@tok@nn}{\let\PYG@bf=\textbf\def\PYG@tc##1{\textcolor[rgb]{0.00,0.00,1.00}{##1}}}
\@namedef{PYG@tok@ne}{\let\PYG@bf=\textbf\def\PYG@tc##1{\textcolor[rgb]{0.82,0.25,0.23}{##1}}}
\@namedef{PYG@tok@nv}{\def\PYG@tc##1{\textcolor[rgb]{0.10,0.09,0.49}{##1}}}
\@namedef{PYG@tok@no}{\def\PYG@tc##1{\textcolor[rgb]{0.53,0.00,0.00}{##1}}}
\@namedef{PYG@tok@nl}{\def\PYG@tc##1{\textcolor[rgb]{0.63,0.63,0.00}{##1}}}
\@namedef{PYG@tok@ni}{\let\PYG@bf=\textbf\def\PYG@tc##1{\textcolor[rgb]{0.60,0.60,0.60}{##1}}}
\@namedef{PYG@tok@na}{\def\PYG@tc##1{\textcolor[rgb]{0.49,0.56,0.16}{##1}}}
\@namedef{PYG@tok@nt}{\let\PYG@bf=\textbf\def\PYG@tc##1{\textcolor[rgb]{0.00,0.50,0.00}{##1}}}
\@namedef{PYG@tok@nd}{\def\PYG@tc##1{\textcolor[rgb]{0.67,0.13,1.00}{##1}}}
\@namedef{PYG@tok@s}{\def\PYG@tc##1{\textcolor[rgb]{0.73,0.13,0.13}{##1}}}
\@namedef{PYG@tok@sd}{\let\PYG@it=\textit\def\PYG@tc##1{\textcolor[rgb]{0.73,0.13,0.13}{##1}}}
\@namedef{PYG@tok@si}{\let\PYG@bf=\textbf\def\PYG@tc##1{\textcolor[rgb]{0.73,0.40,0.53}{##1}}}
\@namedef{PYG@tok@se}{\let\PYG@bf=\textbf\def\PYG@tc##1{\textcolor[rgb]{0.73,0.40,0.13}{##1}}}
\@namedef{PYG@tok@sr}{\def\PYG@tc##1{\textcolor[rgb]{0.73,0.40,0.53}{##1}}}
\@namedef{PYG@tok@ss}{\def\PYG@tc##1{\textcolor[rgb]{0.10,0.09,0.49}{##1}}}
\@namedef{PYG@tok@sx}{\def\PYG@tc##1{\textcolor[rgb]{0.00,0.50,0.00}{##1}}}
\@namedef{PYG@tok@m}{\def\PYG@tc##1{\textcolor[rgb]{0.40,0.40,0.40}{##1}}}
\@namedef{PYG@tok@gh}{\let\PYG@bf=\textbf\def\PYG@tc##1{\textcolor[rgb]{0.00,0.00,0.50}{##1}}}
\@namedef{PYG@tok@gu}{\let\PYG@bf=\textbf\def\PYG@tc##1{\textcolor[rgb]{0.50,0.00,0.50}{##1}}}
\@namedef{PYG@tok@gd}{\def\PYG@tc##1{\textcolor[rgb]{0.63,0.00,0.00}{##1}}}
\@namedef{PYG@tok@gi}{\def\PYG@tc##1{\textcolor[rgb]{0.00,0.63,0.00}{##1}}}
\@namedef{PYG@tok@gr}{\def\PYG@tc##1{\textcolor[rgb]{1.00,0.00,0.00}{##1}}}
\@namedef{PYG@tok@ge}{\let\PYG@it=\textit}
\@namedef{PYG@tok@gs}{\let\PYG@bf=\textbf}
\@namedef{PYG@tok@gp}{\let\PYG@bf=\textbf\def\PYG@tc##1{\textcolor[rgb]{0.00,0.00,0.50}{##1}}}
\@namedef{PYG@tok@go}{\def\PYG@tc##1{\textcolor[rgb]{0.53,0.53,0.53}{##1}}}
\@namedef{PYG@tok@gt}{\def\PYG@tc##1{\textcolor[rgb]{0.00,0.27,0.87}{##1}}}
\@namedef{PYG@tok@err}{\def\PYG@bc##1{{\setlength{\fboxsep}{\string -\fboxrule}\fcolorbox[rgb]{1.00,0.00,0.00}{1,1,1}{\strut ##1}}}}
\@namedef{PYG@tok@kc}{\let\PYG@bf=\textbf\def\PYG@tc##1{\textcolor[rgb]{0.00,0.50,0.00}{##1}}}
\@namedef{PYG@tok@kd}{\let\PYG@bf=\textbf\def\PYG@tc##1{\textcolor[rgb]{0.00,0.50,0.00}{##1}}}
\@namedef{PYG@tok@kn}{\let\PYG@bf=\textbf\def\PYG@tc##1{\textcolor[rgb]{0.00,0.50,0.00}{##1}}}
\@namedef{PYG@tok@kr}{\let\PYG@bf=\textbf\def\PYG@tc##1{\textcolor[rgb]{0.00,0.50,0.00}{##1}}}
\@namedef{PYG@tok@bp}{\def\PYG@tc##1{\textcolor[rgb]{0.00,0.50,0.00}{##1}}}
\@namedef{PYG@tok@fm}{\def\PYG@tc##1{\textcolor[rgb]{0.00,0.00,1.00}{##1}}}
\@namedef{PYG@tok@vc}{\def\PYG@tc##1{\textcolor[rgb]{0.10,0.09,0.49}{##1}}}
\@namedef{PYG@tok@vg}{\def\PYG@tc##1{\textcolor[rgb]{0.10,0.09,0.49}{##1}}}
\@namedef{PYG@tok@vi}{\def\PYG@tc##1{\textcolor[rgb]{0.10,0.09,0.49}{##1}}}
\@namedef{PYG@tok@vm}{\def\PYG@tc##1{\textcolor[rgb]{0.10,0.09,0.49}{##1}}}
\@namedef{PYG@tok@sa}{\def\PYG@tc##1{\textcolor[rgb]{0.73,0.13,0.13}{##1}}}
\@namedef{PYG@tok@sb}{\def\PYG@tc##1{\textcolor[rgb]{0.73,0.13,0.13}{##1}}}
\@namedef{PYG@tok@sc}{\def\PYG@tc##1{\textcolor[rgb]{0.73,0.13,0.13}{##1}}}
\@namedef{PYG@tok@dl}{\def\PYG@tc##1{\textcolor[rgb]{0.73,0.13,0.13}{##1}}}
\@namedef{PYG@tok@s2}{\def\PYG@tc##1{\textcolor[rgb]{0.73,0.13,0.13}{##1}}}
\@namedef{PYG@tok@sh}{\def\PYG@tc##1{\textcolor[rgb]{0.73,0.13,0.13}{##1}}}
\@namedef{PYG@tok@s1}{\def\PYG@tc##1{\textcolor[rgb]{0.73,0.13,0.13}{##1}}}
\@namedef{PYG@tok@mb}{\def\PYG@tc##1{\textcolor[rgb]{0.40,0.40,0.40}{##1}}}
\@namedef{PYG@tok@mf}{\def\PYG@tc##1{\textcolor[rgb]{0.40,0.40,0.40}{##1}}}
\@namedef{PYG@tok@mh}{\def\PYG@tc##1{\textcolor[rgb]{0.40,0.40,0.40}{##1}}}
\@namedef{PYG@tok@mi}{\def\PYG@tc##1{\textcolor[rgb]{0.40,0.40,0.40}{##1}}}
\@namedef{PYG@tok@il}{\def\PYG@tc##1{\textcolor[rgb]{0.40,0.40,0.40}{##1}}}
\@namedef{PYG@tok@mo}{\def\PYG@tc##1{\textcolor[rgb]{0.40,0.40,0.40}{##1}}}
\@namedef{PYG@tok@ch}{\let\PYG@it=\textit\def\PYG@tc##1{\textcolor[rgb]{0.25,0.50,0.50}{##1}}}
\@namedef{PYG@tok@cm}{\let\PYG@it=\textit\def\PYG@tc##1{\textcolor[rgb]{0.25,0.50,0.50}{##1}}}
\@namedef{PYG@tok@cpf}{\let\PYG@it=\textit\def\PYG@tc##1{\textcolor[rgb]{0.25,0.50,0.50}{##1}}}
\@namedef{PYG@tok@c1}{\let\PYG@it=\textit\def\PYG@tc##1{\textcolor[rgb]{0.25,0.50,0.50}{##1}}}
\@namedef{PYG@tok@cs}{\let\PYG@it=\textit\def\PYG@tc##1{\textcolor[rgb]{0.25,0.50,0.50}{##1}}}

\def\PYGZus{\char`\_}

\def\PYGZsh{\char`\#}

\def\PYGZhy{\char`\-}

\makeatother

\makeatletter
\def\PYGdefault@reset{\let\PYGdefault@it=\relax \let\PYGdefault@bf=\relax%
    \let\PYGdefault@ul=\relax \let\PYGdefault@tc=\relax%
    \let\PYGdefault@bc=\relax \let\PYGdefault@ff=\relax}
\def\PYGdefault@tok#1{\csname PYGdefault@tok@#1\endcsname}
\def\PYGdefault@toks#1+{\ifx\relax#1\empty\else%
    \PYGdefault@tok{#1}\expandafter\PYGdefault@toks\fi}
\def\PYGdefault@do#1{\PYGdefault@bc{\PYGdefault@tc{\PYGdefault@ul{%
    \PYGdefault@it{\PYGdefault@bf{\PYGdefault@ff{#1}}}}}}}
\def\PYGdefault#1#2{\PYGdefault@reset\PYGdefault@toks#1+\relax+\PYGdefault@do{#2}}

\@namedef{PYGdefault@tok@w}{\def\PYGdefault@tc##1{\textcolor[rgb]{0.73,0.73,0.73}{##1}}}
\@namedef{PYGdefault@tok@c}{\let\PYGdefault@it=\textit\def\PYGdefault@tc##1{\textcolor[rgb]{0.25,0.50,0.50}{##1}}}
\@namedef{PYGdefault@tok@cp}{\def\PYGdefault@tc##1{\textcolor[rgb]{0.74,0.48,0.00}{##1}}}
\@namedef{PYGdefault@tok@k}{\let\PYGdefault@bf=\textbf\def\PYGdefault@tc##1{\textcolor[rgb]{0.00,0.50,0.00}{##1}}}
\@namedef{PYGdefault@tok@kp}{\def\PYGdefault@tc##1{\textcolor[rgb]{0.00,0.50,0.00}{##1}}}
\@namedef{PYGdefault@tok@kt}{\def\PYGdefault@tc##1{\textcolor[rgb]{0.69,0.00,0.25}{##1}}}
\@namedef{PYGdefault@tok@o}{\def\PYGdefault@tc##1{\textcolor[rgb]{0.40,0.40,0.40}{##1}}}
\@namedef{PYGdefault@tok@ow}{\let\PYGdefault@bf=\textbf\def\PYGdefault@tc##1{\textcolor[rgb]{0.67,0.13,1.00}{##1}}}
\@namedef{PYGdefault@tok@nb}{\def\PYGdefault@tc##1{\textcolor[rgb]{0.00,0.50,0.00}{##1}}}
\@namedef{PYGdefault@tok@nf}{\def\PYGdefault@tc##1{\textcolor[rgb]{0.00,0.00,1.00}{##1}}}
\@namedef{PYGdefault@tok@nc}{\let\PYGdefault@bf=\textbf\def\PYGdefault@tc##1{\textcolor[rgb]{0.00,0.00,1.00}{##1}}}
\@namedef{PYGdefault@tok@nn}{\let\PYGdefault@bf=\textbf\def\PYGdefault@tc##1{\textcolor[rgb]{0.00,0.00,1.00}{##1}}}
\@namedef{PYGdefault@tok@ne}{\let\PYGdefault@bf=\textbf\def\PYGdefault@tc##1{\textcolor[rgb]{0.82,0.25,0.23}{##1}}}
\@namedef{PYGdefault@tok@nv}{\def\PYGdefault@tc##1{\textcolor[rgb]{0.10,0.09,0.49}{##1}}}
\@namedef{PYGdefault@tok@no}{\def\PYGdefault@tc##1{\textcolor[rgb]{0.53,0.00,0.00}{##1}}}
\@namedef{PYGdefault@tok@nl}{\def\PYGdefault@tc##1{\textcolor[rgb]{0.63,0.63,0.00}{##1}}}
\@namedef{PYGdefault@tok@ni}{\let\PYGdefault@bf=\textbf\def\PYGdefault@tc##1{\textcolor[rgb]{0.60,0.60,0.60}{##1}}}
\@namedef{PYGdefault@tok@na}{\def\PYGdefault@tc##1{\textcolor[rgb]{0.49,0.56,0.16}{##1}}}
\@namedef{PYGdefault@tok@nt}{\let\PYGdefault@bf=\textbf\def\PYGdefault@tc##1{\textcolor[rgb]{0.00,0.50,0.00}{##1}}}
\@namedef{PYGdefault@tok@nd}{\def\PYGdefault@tc##1{\textcolor[rgb]{0.67,0.13,1.00}{##1}}}
\@namedef{PYGdefault@tok@s}{\def\PYGdefault@tc##1{\textcolor[rgb]{0.73,0.13,0.13}{##1}}}
\@namedef{PYGdefault@tok@sd}{\let\PYGdefault@it=\textit\def\PYGdefault@tc##1{\textcolor[rgb]{0.73,0.13,0.13}{##1}}}
\@namedef{PYGdefault@tok@si}{\let\PYGdefault@bf=\textbf\def\PYGdefault@tc##1{\textcolor[rgb]{0.73,0.40,0.53}{##1}}}
\@namedef{PYGdefault@tok@se}{\let\PYGdefault@bf=\textbf\def\PYGdefault@tc##1{\textcolor[rgb]{0.73,0.40,0.13}{##1}}}
\@namedef{PYGdefault@tok@sr}{\def\PYGdefault@tc##1{\textcolor[rgb]{0.73,0.40,0.53}{##1}}}
\@namedef{PYGdefault@tok@ss}{\def\PYGdefault@tc##1{\textcolor[rgb]{0.10,0.09,0.49}{##1}}}
\@namedef{PYGdefault@tok@sx}{\def\PYGdefault@tc##1{\textcolor[rgb]{0.00,0.50,0.00}{##1}}}
\@namedef{PYGdefault@tok@m}{\def\PYGdefault@tc##1{\textcolor[rgb]{0.40,0.40,0.40}{##1}}}
\@namedef{PYGdefault@tok@gh}{\let\PYGdefault@bf=\textbf\def\PYGdefault@tc##1{\textcolor[rgb]{0.00,0.00,0.50}{##1}}}
\@namedef{PYGdefault@tok@gu}{\let\PYGdefault@bf=\textbf\def\PYGdefault@tc##1{\textcolor[rgb]{0.50,0.00,0.50}{##1}}}
\@namedef{PYGdefault@tok@gd}{\def\PYGdefault@tc##1{\textcolor[rgb]{0.63,0.00,0.00}{##1}}}
\@namedef{PYGdefault@tok@gi}{\def\PYGdefault@tc##1{\textcolor[rgb]{0.00,0.63,0.00}{##1}}}
\@namedef{PYGdefault@tok@gr}{\def\PYGdefault@tc##1{\textcolor[rgb]{1.00,0.00,0.00}{##1}}}
\@namedef{PYGdefault@tok@ge}{\let\PYGdefault@it=\textit}
\@namedef{PYGdefault@tok@gs}{\let\PYGdefault@bf=\textbf}
\@namedef{PYGdefault@tok@gp}{\let\PYGdefault@bf=\textbf\def\PYGdefault@tc##1{\textcolor[rgb]{0.00,0.00,0.50}{##1}}}
\@namedef{PYGdefault@tok@go}{\def\PYGdefault@tc##1{\textcolor[rgb]{0.53,0.53,0.53}{##1}}}
\@namedef{PYGdefault@tok@gt}{\def\PYGdefault@tc##1{\textcolor[rgb]{0.00,0.27,0.87}{##1}}}
\@namedef{PYGdefault@tok@err}{\def\PYGdefault@bc##1{{\setlength{\fboxsep}{\string -\fboxrule}\fcolorbox[rgb]{1.00,0.00,0.00}{1,1,1}{\strut ##1}}}}
\@namedef{PYGdefault@tok@kc}{\let\PYGdefault@bf=\textbf\def\PYGdefault@tc##1{\textcolor[rgb]{0.00,0.50,0.00}{##1}}}
\@namedef{PYGdefault@tok@kd}{\let\PYGdefault@bf=\textbf\def\PYGdefault@tc##1{\textcolor[rgb]{0.00,0.50,0.00}{##1}}}
\@namedef{PYGdefault@tok@kn}{\let\PYGdefault@bf=\textbf\def\PYGdefault@tc##1{\textcolor[rgb]{0.00,0.50,0.00}{##1}}}
\@namedef{PYGdefault@tok@kr}{\let\PYGdefault@bf=\textbf\def\PYGdefault@tc##1{\textcolor[rgb]{0.00,0.50,0.00}{##1}}}
\@namedef{PYGdefault@tok@bp}{\def\PYGdefault@tc##1{\textcolor[rgb]{0.00,0.50,0.00}{##1}}}
\@namedef{PYGdefault@tok@fm}{\def\PYGdefault@tc##1{\textcolor[rgb]{0.00,0.00,1.00}{##1}}}
\@namedef{PYGdefault@tok@vc}{\def\PYGdefault@tc##1{\textcolor[rgb]{0.10,0.09,0.49}{##1}}}
\@namedef{PYGdefault@tok@vg}{\def\PYGdefault@tc##1{\textcolor[rgb]{0.10,0.09,0.49}{##1}}}
\@namedef{PYGdefault@tok@vi}{\def\PYGdefault@tc##1{\textcolor[rgb]{0.10,0.09,0.49}{##1}}}
\@namedef{PYGdefault@tok@vm}{\def\PYGdefault@tc##1{\textcolor[rgb]{0.10,0.09,0.49}{##1}}}
\@namedef{PYGdefault@tok@sa}{\def\PYGdefault@tc##1{\textcolor[rgb]{0.73,0.13,0.13}{##1}}}
\@namedef{PYGdefault@tok@sb}{\def\PYGdefault@tc##1{\textcolor[rgb]{0.73,0.13,0.13}{##1}}}
\@namedef{PYGdefault@tok@sc}{\def\PYGdefault@tc##1{\textcolor[rgb]{0.73,0.13,0.13}{##1}}}
\@namedef{PYGdefault@tok@dl}{\def\PYGdefault@tc##1{\textcolor[rgb]{0.73,0.13,0.13}{##1}}}
\@namedef{PYGdefault@tok@s2}{\def\PYGdefault@tc##1{\textcolor[rgb]{0.73,0.13,0.13}{##1}}}
\@namedef{PYGdefault@tok@sh}{\def\PYGdefault@tc##1{\textcolor[rgb]{0.73,0.13,0.13}{##1}}}
\@namedef{PYGdefault@tok@s1}{\def\PYGdefault@tc##1{\textcolor[rgb]{0.73,0.13,0.13}{##1}}}
\@namedef{PYGdefault@tok@mb}{\def\PYGdefault@tc##1{\textcolor[rgb]{0.40,0.40,0.40}{##1}}}
\@namedef{PYGdefault@tok@mf}{\def\PYGdefault@tc##1{\textcolor[rgb]{0.40,0.40,0.40}{##1}}}
\@namedef{PYGdefault@tok@mh}{\def\PYGdefault@tc##1{\textcolor[rgb]{0.40,0.40,0.40}{##1}}}
\@namedef{PYGdefault@tok@mi}{\def\PYGdefault@tc##1{\textcolor[rgb]{0.40,0.40,0.40}{##1}}}
\@namedef{PYGdefault@tok@il}{\def\PYGdefault@tc##1{\textcolor[rgb]{0.40,0.40,0.40}{##1}}}
\@namedef{PYGdefault@tok@mo}{\def\PYGdefault@tc##1{\textcolor[rgb]{0.40,0.40,0.40}{##1}}}
\@namedef{PYGdefault@tok@ch}{\let\PYGdefault@it=\textit\def\PYGdefault@tc##1{\textcolor[rgb]{0.25,0.50,0.50}{##1}}}
\@namedef{PYGdefault@tok@cm}{\let\PYGdefault@it=\textit\def\PYGdefault@tc##1{\textcolor[rgb]{0.25,0.50,0.50}{##1}}}
\@namedef{PYGdefault@tok@cpf}{\let\PYGdefault@it=\textit\def\PYGdefault@tc##1{\textcolor[rgb]{0.25,0.50,0.50}{##1}}}
\@namedef{PYGdefault@tok@c1}{\let\PYGdefault@it=\textit\def\PYGdefault@tc##1{\textcolor[rgb]{0.25,0.50,0.50}{##1}}}
\@namedef{PYGdefault@tok@cs}{\let\PYGdefault@it=\textit\def\PYGdefault@tc##1{\textcolor[rgb]{0.25,0.50,0.50}{##1}}}

\makeatother

\title{Spectral decoupling for training transferable neural networks in medical imaging}

\usepackage{authblk}

\author[1]{Joona Pohjonen\protect\textsuperscript{*}}
\author[1]{Carolin Stürenberg}
\author[1,2]{Antti Rannikko\protect\textsuperscript{$\dagger$}}
\author[1,3]{Tuomas Mirtti\protect\textsuperscript{$\dagger$}}
\author[4,5]{Esa Pitkänen\protect\textsuperscript{*$\dagger$}}

\affil[1]{Research Program in Systems Oncology, University of Helsinki}
\affil[2]{Department of Urology, Helsinki University Hospital}
\affil[3]{Department of Pathology, Helsinki University Hospital}
\affil[4]{Institute for Molecular Medicine Finland (FIMM), HiLIFE, University of Helsinki}
\affil[5]{Research Program in Applied Tumor Genomics, University of Helsinki}

\begin{document}

\twocolumn[
    \begin{@twocolumnfalse}

    \maketitle
    \vspace{0.4cm}

    \begin{abstract}
    Many current neural networks for medical imaging generalise poorly to data unseen during training. Such behaviour can be caused by networks overfitting easy-to-learn, or statistically dominant, features while disregarding other potentially informative features. For example, indistinguishable differences in the sharpness of the images from two different scanners can degrade the performance of the network significantly. All neural networks intended for clinical practice need to be robust to variation in data caused by differences in imaging equipment, sample preparation and patient populations.
    
    To address these challenges, we evaluate the utility of spectral decoupling as an implicit bias mitigation method. Spectral decoupling encourages the neural network to learn more features by simply regularising the networks' unnormalised prediction scores with an L2 penalty, thus having no added computational costs.
    
    We show that spectral decoupling allows training neural networks on datasets with strong spurious correlations and increases networks' robustness for data distribution shifts. To validate our findings, we train networks with and without spectral decoupling to detect prostate cancer tissue slides and COVID-19 in chest radiographs. Networks trained with spectral decoupling achieve up to 9.5 percent point higher performance on external datasets.
    
    Our results show that spectral decoupling helps with generalisation issues associated with neural networks, and can be used to complement or replace computationally expensive explicit bias mitigation methods, such as stain normalization in histological images. We recommend using spectral decoupling as an implicit bias mitigation method in any neural network intended for clinical use.
    \end{abstract}
    
    \vspace{0.8cm}
    \end{@twocolumnfalse}
]

{
  \renewcommand{\thefootnote}{*}
  \footnotetext[1]{
    Correspondence:\\ \hspace*{2.5em}\texttt{joona.pohjonen@helsinki.fi} (lead contact)\\
    \hspace*{2.5em}\texttt{esa.pitkanen@helsinki.fi}
  }
  \renewcommand{\thefootnote}{$\dagger$}
  \footnotetext[2]{
    Senior author
  }
}



\section{Introduction}

Neural networks have been adapted to many medical imaging tasks with impressive results, often surpassing human counterparts in consistency, speed and accuracy \cite{liu2019comparison}. However, these networks are prone to overfit easy-to-learn, or statistically dominant, features while disregarding other potentially informative features. This leads to poor generalisation to data generated by different medical centres, reliance on the dominant features, and lack of robustness \cite{pezeshki2020sd, geirhos2020shortcut}. For example, a neural network classifier for skin cancer, approved to be used as a medical device in Europe, had overfit the correlation between surgical margins and malignant melanoma \cite{winkler2019melanoma}. Due to this, the false positive rate of the network was increased by 40 percentage points during external validation. Furthermore, three out of five neural networks for pneumonia detection showed significantly worse performance during external validation \cite{zech2018pneumonia} and recent neural networks for COVID-19 detection rely on confounding factors rather than actual medical pathology \cite{degrave2021covid}. Even small differences in the sharpness of images from two different scanners can degrade the performance of neural networks significantly (see Section \ref{sec:robustness}).

Although generalisation issues need to be solved before any neural networks can be applied in clinical practice, the phenomenon is still poorly understood \cite{van2021toclinic}. This may be because the detection of generalisation issues is hard and often requires state-of-the-art methods of explainable AI \cite{degrave2021covid}. An external dataset is one of the only methods of testing generalization performance, although it will uncover generalisation issues only when the neural network fails to generalize to the dataset. If a neural network achieves high overall accuracy on the external dataset, it may still always fail for some subset of samples. Any particular external dataset may also contain the same sources of bias as the training data.

Explicit methods have been proposed to address specific sources of bias, like using augmentation to address staining differences in tissue section slides \cite{tellez2019quantifying} or normalising each image with a common standard \cite{bel2019normalize, janowczyk2017stain}. The obvious problem with explicit methods is that they only control for selected biases and more subtle sources of bias, like small differences between patient populations, may go unaddressed. Implicit methods of bias control are required before neural networks can be safely applied to clinical practice.

Learning dominant features at the cost of other potentially informative features, also known as shortcut-learning, is a common problem in all neural networks and one of the main reasons behind the generalisation issues \cite{geirhos2020shortcut}. Shortcut-learning occurs mainly because of gradient starvation, where gradient descent updates the parameters of a neural network in directions capturing only dominant features, thus starving the gradient from other features \cite{combes2018gs}. The gradient descent algorithm finds a local optimum by taking small steps towards the opposite sign of the derivative, the direction of the steepest descent \cite{cauchy1847gradient_descent}. The recently proposed method of spectral decoupling \cite{pezeshki2020sd} provably decouples the learning dynamics leading to gradient starvation when using cross-entropy loss, thus encouraging the network to learn more features. The effect is achieved by simply adding an L2 penalty on the unnormalised prediction scores (logits) of the network.

We evaluate the utility of spectral decoupling as an implicit bias mitigation method in the context of medical imaging. We use simulation experiments to show that spectral decoupling increases networks' robustness to data distribution shifts and can be used to train generalisable networks on datasets with a strong superficial correlation. The findings are then evaluated by training prostate cancer and COVID-19 classifiers, where the networks trained with spectral decoupling achieve significantly higher performance on all evaluation datasets.


\section{Materials and methods}

\subsection{Spectral decoupling}
\label{sec:sd}

In spectral decoupling, the network is regularised by imposing an L2 penalty on the unnormalised outputs of the last layer of the network, or logits $\hat{y}$, which is then added to cross-entropy loss, $\mathcal{L}_{\text{CE}}$. This penalty provably \cite{pezeshki2020sd} avoids the conditions leading to gradient starvation in networks trained with cross-entropy loss. Two variants of the penalty are defined as

\begin{equation}
    \mathcal{L}_{\text{CE}} + \frac{\lambda}{2}||\hat{y}||^2_2,
    \label{eq:simple}
\end{equation}

\begin{equation}
    \mathcal{L}_{\text{CE}} + \frac{\lambda}{2}||\hat{y}-\gamma||^2_2.
    \label{eq:complex}
\end{equation}

For Equation \ref{eq:simple}, there is a single tunable hyper-parameter $\lambda$. For Equation \ref{eq:complex}, hyper-parameters $\lambda$ and $\gamma$ are tuned separately for each class, a total of four hyper-parameters for the binary classification task in our study. Pseudo-code for implementing Equation \ref{eq:simple} is presented in Algorithm \ref{alg:eq1}.

\begin{algorithm}[t]
\caption{PyTorch style pseudo-code for Equation \ref{eq:simple}}
\label{alg:eq1}
\begin{Verbatim}[commandchars=\\\{\}]
\PYG{c+c1}{\PYGZsh{} net:       Neural network}
\PYG{c+c1}{\PYGZsh{} lambda\PYGZus{}sd: Hyper\PYGZhy{}parameter for Equation 1.}

\PYG{k}{for} \PYG{p}{(}\PYG{n}{images}\PYG{p}{,} \PYG{n}{targets}\PYG{p}{)} \PYG{o+ow}{in} \PYG{n}{loader}\PYG{p}{:}
    \PYG{c+c1}{\PYGZsh{} Pass images through the network.}
    \PYG{n}{logits} \PYG{o}{=} \PYG{n}{net}\PYG{p}{(}\PYG{n}{images}\PYG{p}{)}
    \PYG{c+c1}{\PYGZsh{} Compute cross\PYGZhy{}entropy loss.}
    \PYG{n}{loss} \PYG{o}{=} \PYG{n}{cross\PYGZus{}entropy}\PYG{p}{(}\PYG{n}{logits}\PYG{p}{,} \PYG{n}{targets}\PYG{p}{)}
    \PYG{c+c1}{\PYGZsh{} Add spectral decoupling penalty.}
    \PYG{n}{loss} \PYG{o}{+=} \PYG{n}{lambda\PYGZus{}sd}\PYG{o}{/}\PYG{l+m+mi}{2} \PYG{o}{*} \PYG{p}{(}\PYG{n}{logits}\PYG{o}{**}\PYG{l+m+mi}{2}\PYG{p}{)}\PYG{o}{.}\PYG{n}{mean}\PYG{p}{()}
    \PYG{c+c1}{\PYGZsh{} Optimization step.}
    \PYG{n}{loss}\PYG{o}{.}\PYG{n}{backward}\PYG{p}{()}
    \PYG{n}{optimizer}\PYG{o}{.}\PYG{n}{step}\PYG{p}{()}
\end{Verbatim}
%
\end{algorithm}

A simple grid search is used to optimize the hyper-parameters in Sections \ref{sec:robustness}, \ref{sec:real_world} and \ref{sec:covid_detection}. Bayesian optimisation is used in Section \ref{sec:features}. Search spaces for the grid search are defined as $S_1 = \{0.1, 0.01, \dots, 0.000001\}$, $S_2 = \{-1, 0, 1, 2\}$, where $\lambda, \lambda_{pos}, \lambda_{neg} \in S_1$ and $\gamma_{pos},\gamma_{pos} \in S_2$. Hyper-parameter optimization is done on the validation split, except for Equation \ref{eq:complex} in Section \ref{sec:features}, where we perform optimization straight on the test split. For Equation \ref{eq:simple}, the tuned hyper-parameter is $\lambda=0.01$. For Equation \ref{eq:complex}, the tuned hyper-parameters are $\lambda_{neg}=0.0969$, $\gamma_{neg}=1.83$, $\lambda_{pos} = 0.000698$ and $\gamma_{pos} = 2.61$ for the experiment in Section \ref{sec:features}, and $\lambda_{neg}=0.01$, $\gamma_{neg}=0$, $\lambda_{pos} = 0.001$ and $\gamma_{pos} = 1$ for the experiment in Section \ref{sec:covid_detection}.

\subsection{Datasets}
\label{dataset:prostate}

\subsubsection{Prostate}

A total of 30 prostate cancer patient cases are annotated for classification into cancerous and benign tissue, where the cancerous areas were annotated in consensus by two observers (C.S., T.M.). All patients have undergone radical prostatectomy at the Helsinki University Hospital between 2014 and 2015. Each case contains 14 to 21 tissue section slides of the prostate. Tissue sections have a thickness of 4 $\mu$m and were stained with hematoxylin and eosin in a clinical-grade laboratory at the Helsinki University Hospital Diagnostic Center, Department of Pathology. Two different scanners are used to obtain images of the tissue section slides at 20x magnification. Larger macro slides (whole-mount, 2x3 inch slides) are scanned with an Axio Scan Z.1 scanner (Zeiss, Oberkochen, Germany), and the normal size slides with a Pannoramic Flash III 250 scanner (3DHistech, Budapest, Hungary). From the 30 patient cases, five are set aside for a test set and four are used as a validation set during training and hyper-parameter tuning. The test set is further divided based on the scanner used to obtain the images. Digital slide images are cut into tiles with $1024 \times 1024$ pixels and 20\% overlap, resulting in 4.7 million tiles with 10\% containing cancerous tissue.

To test the differences between cohorts from the same medical centre, another set of 60 prostate cancer patient cases are annotated into cancerous and benign tissue by one of six experienced pathologists. All patients have undergone radical prostatectomy at the Helsinki University Hospital between 2019 and 2020. Each case contains 10 to 21 normal and macro tissue section slides of the prostate. Tissue sections have a thickness of 4 $\mu$m and are also stained with hematoxylin and eosin in a clinical-grade laboratory at the Helsinki University Hospital Diagnostic Center, Department of Pathology. All slides are scanned with an Axio Scan Z.1 scanner (Zeiss, Oberkochen, Germany). Digital slide images are cut into tiles with $1024 \times 1024$ pixels and 20\% overlap, resulting in 13.1 million tiles with 16\% containing cancerous tissue.

For external validation, a freely available prostate cancer dataset is used, containing tissue section slides from patients who have undergone a radical prostatectomy at the Radboud University Medical Center between 2006 and 2011 \cite{PESO_dataset, bulten2019epithelium}. The dataset contains images with $2500 \times 2500$ pixels annotated by a uropathologist as either cancerous or benign. Images are scanned with a Pannoramic Flash II 250 scanner (3DHistech, Budapest, Hungary) at 20x magnification but later reduced to 10x magnification. These images are cut into tiles with $512 \times 512$ pixels and 20\% overlap, resulting in 5655 tiles with 45\% containing cancerous tissue.

All digital slide images are cut and processed with \texttt{HistoPrep} \cite{pohjonen2021hp}. A summary of the prostate datasets is presented in Table \ref{tab:datasets}.

\begin{table*}[t]
  \centering
  \caption{Prostate datasets}
  \begin{tabular}{p{0.225\textwidth}llllll}
    \toprule
    Centre & Scanner & Slides & Tiles & Train data & Test data \\
    \midrule
    \multirow[t]{3}{*}{Helsinki University Hospital} & Pannoramic Flash III 250 & Normal & 1.0 million & Section \ref{sec:features} &  Section \ref{sec:real_world}\\
        & Axio Scan Z.1 & Macro & 3.7 million & Section \ref{sec:robustness} \& \ref{sec:real_world} &  Section \ref{sec:robustness} \& \ref{sec:real_world}\\ 
        & Axio Scan Z.1 & Macro & 13.1 million & – &  Section \ref{sec:real_world} \\ 
    Radboud University Medical Center & Pannoramic Flash II 250 & Both & 5655 & – & Section \ref{sec:real_world} \\
    \bottomrule
  \end{tabular}
  \label{tab:datasets}
\end{table*}

\subsubsection{COVID-19}
\label{dataset:covid}

For COVID-19 detection, we use large open-access repositories of chest radiographs. COVIDx8 dataset is compiled from five different open-source repositories and contains radiographs from over 15,000 patient cases from at least 51 countries, with over 1500 COVID-19 positive patient cases \cite{cohen2020covid, Wang2020covidnet, tsai2021rsna, rahman2021exploring, chowdhury2020data}. BIMCV$\pm$ dataset (iteration 2) contains 3033 positive and 2743 negative COVID-19 patient cases, and 9171 radiographs, after exclusions, collected from the multiple same medical centres during the same time period \cite{de2020bimcv}. Only PA and upright AP radiographs \cite{cohen2020covid} with windowing information were selected from the BIMCV$\pm$ dataset. PadChest dataset contains over 67,000 COVID-19 negative patient cases, and 114,227 radiographs from a single medical centre in Valencia, Spain \cite{bustos2020padchest}. 19 corrupted images were excluded from the PadChest dataset.

COVIDx8 dataset is reserved as an external dataset, and two training datasets are compiled by using only the BIMCV$\pm$ dataset and by adding the PadChest and BIMCV$\pm$ datasets together. 5\% of both training datasets are set aside for validation.

\subsection{Simulation datasets}

Two simulation experiments are used to more closely investigate the utility of spectral decoupling as an implicit bias mitigation method. For both experiments, the dataset from Helsinki University Hospital described in Section \ref{dataset:prostate} is modified in specific ways.

\subsubsection{Cutout dataset}
\label{dataset:sim1}

\begin{figure}[t]
  \centering
  \includegraphics[width=0.9\columnwidth]{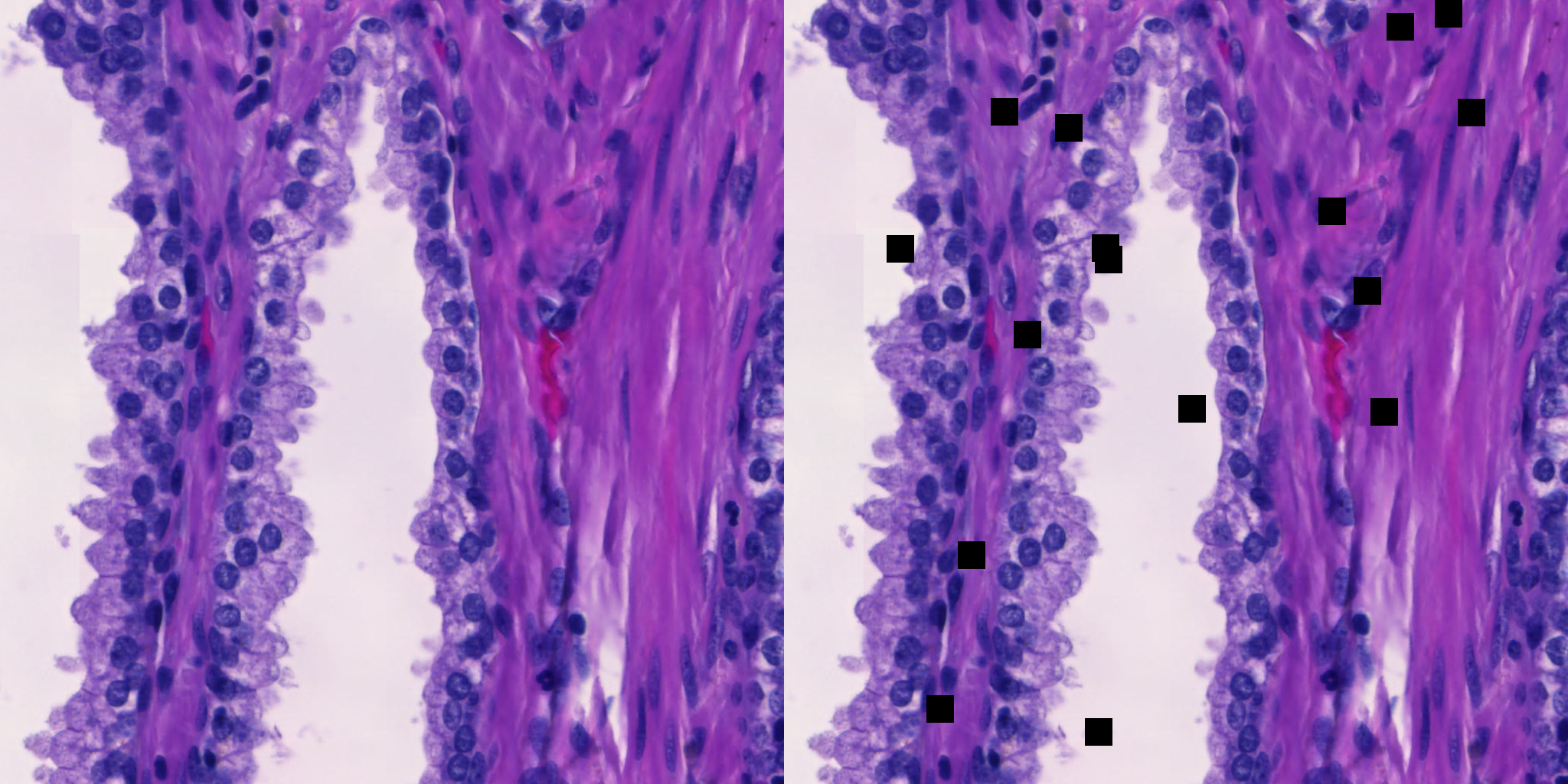}
  \caption{Left: Benign sample. Right: 16 cutouts of $8 \times 8$ pixels added to the benign sample.}
  \label{fig:cutout}
\end{figure}

A dominant feature present in a real-world dataset could be, for example, a biological marker, a certain cancer type or a scanner artefact. To represent these kinds of features, 16 cutouts of $8 \times 8$ pixels are added to the images (Figure \ref{fig:cutout}).

For the experiment, 200,000 images are selected for the training set with an equal amount of samples with cancerous and benign annotations. For the training set, cutouts are added to 25\% and 2.5\% of the benign and cancerous samples, respectively. This makes the presence of cutouts in the image spuriously correlated with a benign annotation. If the network overfits this correlation, cancerous samples with cutouts may be classified as benign. Thus for the test set, cutouts are added to all cancerous samples and none of the benign samples. For a control training set, cutouts are added to all images. Networks trained with this dataset provide a reference point of the performance with cutouts but without the spurious correlation.

\subsubsection{Robustness dataset}
\label{dataset:sim2}

Shifts from the training data distribution are common when evaluating the neural network with datasets from different medical centres. Small changes in the images due to differences in, for example, sample preparation or imaging equipment can cause shifts from the training data distribution. We assess the networks' robustness to these data distribution shifts, by applying transformations with increasing magnitudes to the images in the test set. Image sharpness and stain intensity were selected to represent possible dataset shifts caused by differences in the used scanner and sample preparation, respectively. 

The \texttt{UniformAugment} augmentation strategy consists of applying random transformations with a uniformly sampled magnitude to the images before feeding them to the network \cite{lingchen2020uniformaug}. Sharpening the image is included in the set of possible transformations \cite{cubuk2019autoaugment}, meaning that the network sees sharpened images during training. Thus, the data distribution shift caused by sharpening images is being explicitly mitigated, which should help the network to predict correct labels for evaluation images with higher sharpness. Blurring the image is not included in the set of possible transformations \cite{cubuk2019autoaugment}, meaning that the network will not see randomly blurred images during training. Thus, the data distribution shift caused by blurring the images will not be explicitly mitigated and the use of \texttt{UniformAugment} should not directly help the network with blurry evaluation images.

\begin{figure}[t]
  \centering
  \includegraphics[width=\columnwidth]{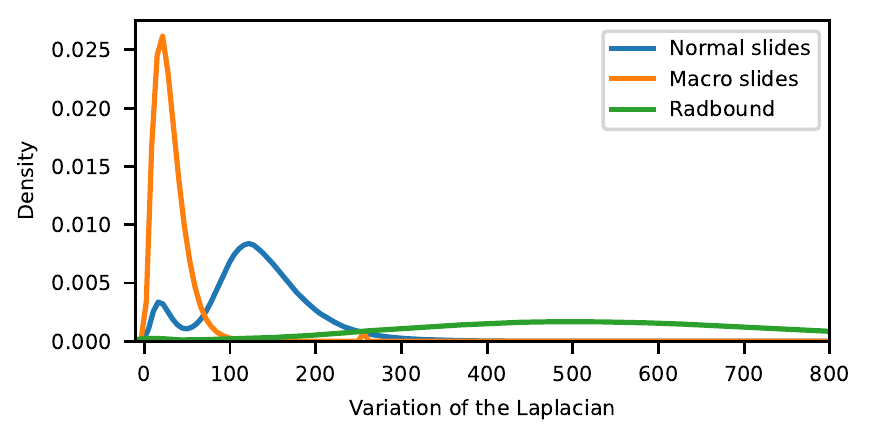}
  \caption{Kernel density estimation of the variance of the images after a Laplace transformation. A higher variance indicates a sharper image. The image is generated from the pre-processing metrics calculated by \texttt{HistoPrep} \cite{pohjonen2021hp}.}
  \label{fig:lap_var}
\end{figure}

By evaluating the network with increasingly sharpened or blurred images, it is possible to assess whether spectral decoupling can improve upon situations where the data distribution shift is, and is not explicitly addressed. Additionally, there are large differences in the sharpness values of real-world datasets from different medical centres and scanners (Figure \ref{fig:lap_var}).

Step-wise blurring is achieved by simple averaging with a $n \times n$ kernel, where $n \in \{2,\dots,20\}$. Sharpened version of the image $x_{\text{sharp}}$ is created by applying kernel
$$
\begin{bmatrix}
    -1 & -1 & -1\\
    -1 &  9 & -1\\
    -1 & -1 & -1\\
\end{bmatrix}
$$
to the original image $x_{\text{original}}$. Sharpness is then gradually increased by creating a new image $x_{\text{blend}}$ with
$$
x_{\text{blend}} = (1-\alpha) x_{\text{original}} + \alpha x_{\text{sharp}},
$$
where $\alpha \in \{0,0.1,\dots,1\}$ defines the amount of sharpness increase.

\begin{figure}[t]
  \centering
  \includegraphics[width=\columnwidth]{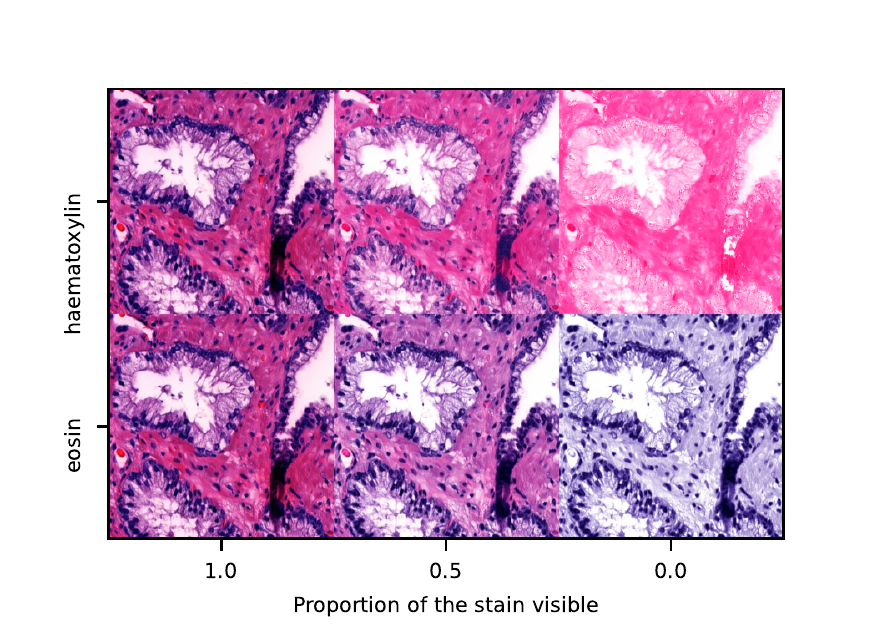}
  \caption{Separation of the heamatoxylin and eosin stains with the Macenko method.}
  \label{fig:macenko}
\end{figure}

To assess the data distribution shifts caused by differences in sample preparation, the intensity of haematoxylin and eosin stains are computationally modified. Haematoxylin highlights cell nuclei, and eosin cytoplasm, connective tissue and muscle. The stain intensities depend on multiple steps in the staining process, and thus the final colour distribution of the slide images varies a lot \cite{tellez2019quantifying}. The stain intensity modification is achieved by first separating the haematoxylin and eosin stains with the Macenko method \cite{macenko2009}. The concentrations of each stain can then be reduced by multiplication with a value between 0 and 1 before the stains are combined back together. An example of the method is shown in Figure \ref{fig:macenko}.

\subsection{Training details}
\label{sec:train}

EfficientNet-b0 network \cite{tan2019efficientnet}, with dropout \cite{srivastava2014dropout} and stochastic depth \cite{huang2016stochastic_depth} of 20\% and an input size of $224 \times 224$, is used as a prostate cancer classifier for all experiments. For augmentation, the input images are randomly cropped and flipped, resized, and then transformed with \texttt{UniformAugment} \cite{lingchen2020uniformaug}, using a maximum of two transformations. Each network is trained for 90 epochs, with a learning rate of $0.005\frac{\text{batch size}}{512}$ and cosine scheduling. Weight decay of 0.0001 is used for networks trained without spectral decoupling. When training neural networks with spectral decoupling, weight decay is disabled.

For COVID-19 detection, we replicate the training regimen from \cite{degrave2021covid}, where a DenseNet-121 network \cite{huang2018densely} is pre-trained with the ImageNet dataset and then fine-tuned for 30 epochs as a binary COVID-19 classifier. All hyper-parameters, other than spectral decoupling, are set to values reported in the paper.

For spectral decoupling, Equation \ref{eq:complex} is used for the first simulation experiment on dominant features (Section \ref{sec:features}) and COVID-19 detection (Section \ref{sec:covid_detection}). Equation \ref{eq:simple} is used for all other experiments (Sections \ref{sec:robustness} and \ref{sec:real_world}).

Each experiment is repeated five times and the summary metrics for these runs are reported. All reported performance metrics are balanced between the classes when necessary and a cut-off value of 0.5 is used to obtain a binary label from the normalised predictions of the network. To compare paired receiver under the operating characteristic (ROC) curves, we use one-tailed DeLong's test and report the $Z$-values and p-values \cite{delong1988auc_test}.

PyTorch (version 1.8) \cite{pytorch} is used for training the neural networks, \texttt{timm} (version 0.1.8) \cite{rw2019timm} for building the neural networks and \texttt{albumentations} (version 0.5.1) \cite{albumentations} for image augmentations.


\section{Experiments}

In this section, the utility of using spectral decoupling as an implicit bias mitigation method is explored with both simulation and real-world experiments.

\subsection{Dominant features}
\label{sec:features}

To assess the utility of spectral decoupling in situations where the training dataset contains a strong dominant feature, the cutout dataset defined in Section \ref{dataset:sim1} is used. Five networks are trained with either spectral decoupling or weight decay on the training set. Additionally, five networks are trained on the control dataset with weight decay to provide a reference point of the performance under no spurious correlation caused by the dominant feature. The mean and standard deviation of the accuracy and recall metrics on the test data are reported in Table \ref{tab:sim_results}. Accuracy is defined as the fraction of all instances that were correctly identified, and recall as the fraction of positive instances that were correctly identified. 

\begin{table}[t]
  \centering
  \caption{Results of the simulation study with the cutout dataset on dominant features. The mean and standard deviation (SD) values are reported for each set of five trained networks.}
  \begin{tabular}{lll}
    \toprule
    Name                   & Accuracy (SD)   & Recall (SD)       \\
    \midrule
    Weight decay           & 0.752 (0.019)    & 0.523 (0.039)    \\
    Spectral decoupling    & 0.837 (0.020)    & 0.715 (0.046)    \\
    Control + weight decay & 0.875 (0.009)    & 0.832 (0.036)    \\
    \bottomrule
  \end{tabular}
  \label{tab:sim_results}
\end{table}

The use of spectral decoupling increases the accuracy by 8.5 percentage points over weight decay and almost reaches the performance of the network trained on the control dataset. The networks trained without spectral decoupling appear to make false predictions based on the dominant feature, although the class activation maps \cite{gradcam2018} of the trained neural networks, do not significantly differ between weight decay and spectral decoupling. As hyper-parameters were tuned on the test set, the results should be interpreted only as a demonstration that spectral decoupling can offer an important level of control over the features that are learned. 

The simpler variant of spectral decoupling in Equation \ref{eq:simple} did not increase the networks' performance in any way, and only after extensive hyper-parameter tuning, Equation \ref{eq:complex} produced the reported results. The hyper-parameter tuning was sensitive to the selected parameters, and even small changes to the final values significantly reduced the accuracy of the neural network. Similar results were also reported with the real-world example in the original paper \cite{pezeshki2020sd}. As extensive hyper-parameter tuning can deter researchers from applying the method, we limit hyper-parameter tuning to a simple grid search over limited search spaces for all other experiments, as described in Section \ref{sec:sd}.

\subsection{Robustness}
\label{sec:robustness}

\begin{figure}[t]
  \centering
  \includegraphics[width=\columnwidth]{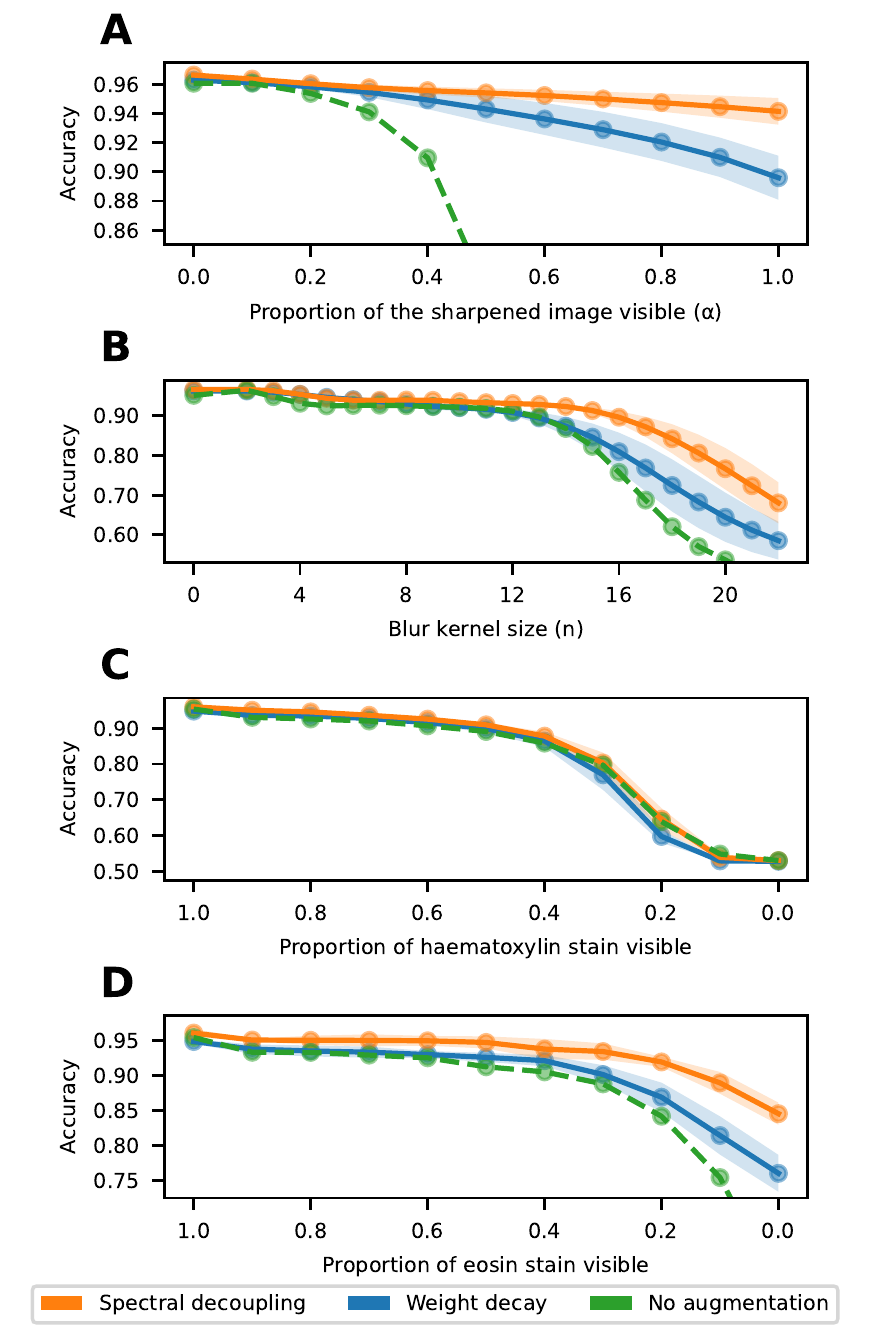}
  \caption{Robustness for data distribution shifts from the training data. The lines show the mean accuracy and the shaded regions represent one standard deviation around the mean.}
  \label{fig:robustness}
\end{figure}

To assess whether spectral decoupling increases neural networks' robustness to data distribution shifts, five networks are trained with either spectral decoupling or weight decay and evaluated on the robustness dataset described in Section \ref{dataset:sim2}. Additionally, five networks are trained with weight decay but without \texttt{UniformAugment} to assess how much the augmentation strategy improves robustness. The robustness to data distribution shifts caused by sharpening, blurring and reducing the intensity of either haematoxylin or eosin stain are presented in Figure \ref{fig:robustness}.

Performance of all networks trained with weight decay and without the augmentation strategy degrade to roughly 50\% accuracy. Training the networks again with \texttt{UniformAugment} significantly increases robustness to all data distribution shifts except with haematoxylin stain intensity reduction (Figure \ref{fig:robustness}C). When the data distribution shift is included as a possible augmentation (Figure \ref{fig:robustness}A), the increase in accuracy is almost 40 percentage points with the most severe distribution shift. When the data distribution shift is not included as a possible transformation (Figure \ref{fig:robustness}B–D), robustness is more similar with and without augmentation. This result demonstrates the importance of using augmentation as an explicit bias mitigation method.

Although the use of augmentation already increased the accuracy by almost 40 percentage points, the use of spectral decoupling is able to improve the accuracy by a further 4.6 percentage points with the most severe data distribution shift (Figure \ref{fig:robustness}A). The increase in accuracy is more pronounced with blurring, 12.4 percentage points with $n=19$ (Figure \ref{fig:robustness}B), and eosin stain intensity reduction, where networks trained with spectral decoupling achieve 1.2 to 8.5 percentage points higher accuracy with a 0.9 to 0.0 multiplier (Figure \ref{fig:robustness}D). These data distribution shifts are not included as a possible transformations in \texttt{UniformAugment}, and thus not explicitly controlled. With haematoxylin stain intensity reduction, all networks degrade similarly in performance (Figure \ref{fig:robustness}C). These results show that spectral decoupling is able to significantly complement and improve upon augmentation, as well as improve robustness to data distribution shifts that are not explicitly controlled by augmentation.

\subsection{Prostate cancer detection}
\label{sec:real_world}

To assess whether the results of the simulation experiments translate into improvements in real-world datasets, we train networks with and without spectral decoupling to detect prostate cancer on haematoxylin and eosin stained whole slide images of the prostate. These networks are then evaluated on four different datasets described in Section \ref{dataset:prostate}.

\begin{figure}[t]
    \centering
    \includegraphics[width=\columnwidth]{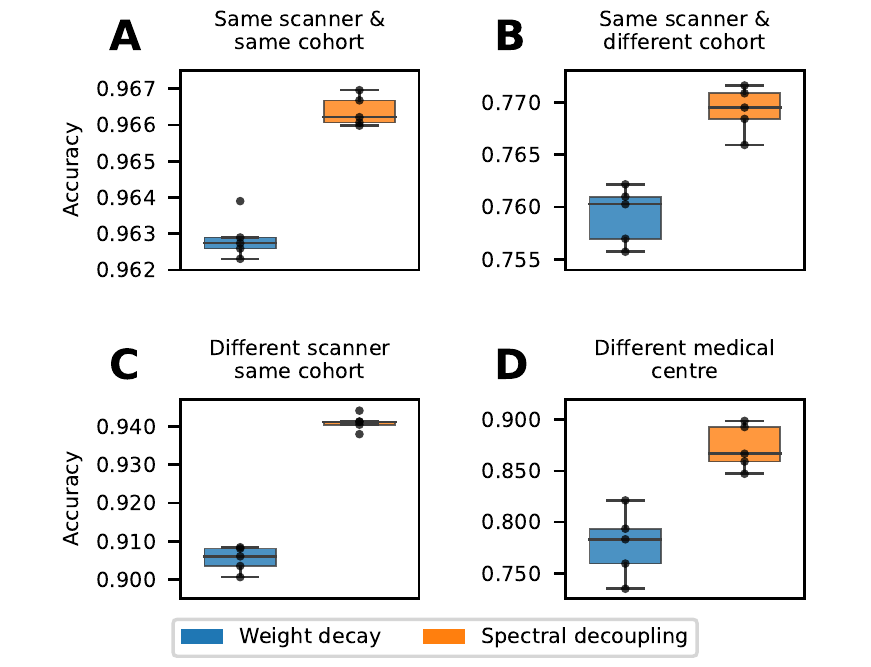}
    \caption{Neural network performance on evaluation datasets. Each consecutive evaluation dataset moves further from the training data distribution. Networks trained with spectral decoupling improve accuracy by 0.35 (\textbf{A}), 1.0 (\textbf{B}), 3.6 (\textbf{C}) and 9.5 (\textbf{D}) percentage points over weight decay. All models are trained with \texttt{UniformAugment}.}
    \label{fig:real_world}
\end{figure}

The results are presented in Figure \ref{fig:real_world}. Networks trained with spectral decoupling show higher performance on all evaluation datasets. The difference between weight decay and spectral decoupling gets more pronounced as we move further away from the training dataset distribution. Finally, there is a 9.5 percentage point increase in accuracy over weight decay on the dataset from a different medical centre. The reported performances are not comparable between evaluation datasets, as each dataset has been annotated with a different strategy and thus contain different amounts of label noise.

\begin{figure}[t]
  \centering
  \includegraphics[width=\columnwidth]{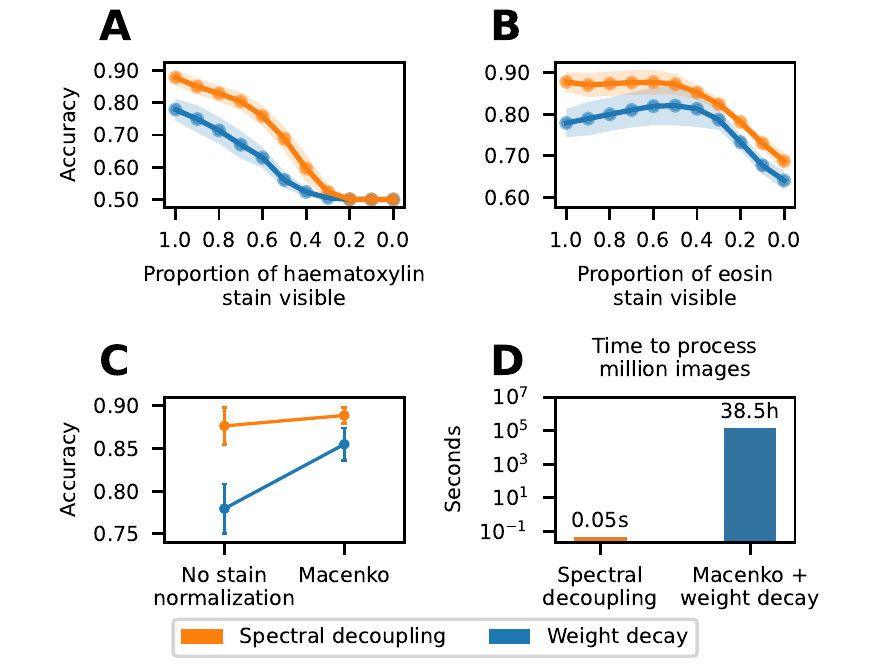}
   \caption{Spectral decoupling can complement or even replace computationally heavy stain normalization methods. Robustness to data distribution shifts, on the external dataset, caused by haematoxylin (\textbf{A}) or eosin (\textbf{B}) stain intensity reduction. (\textbf{C}) Network accuracy increases when normalizing haematoxylin and eosin stain intensities with the Macenko method. (\textbf{D}) Comparison of the computational requirements between spectral decoupling and the Macenko method. Images per seconds estimation for spectral decoupling is calculated with a Equation \ref{eq:simple}, where $\hat{y}$ is a $512 \times 1$ matrix and Macenko stain normalisation is performed on resized images of size $224 \times 224$.}
  \label{fig:peso}
\end{figure}

To further explore why networks trained without spectral decoupling fail to generalise to the dataset from Radboud University Medical Center (Figure \ref{fig:real_world}D), the robustness to haematoxylin and eosin stain intensities are explored in Figures \ref{fig:peso}A–B. Spectral decoupling is less sensitive to both haematoxylin and eosin stain intensity reduction and interestingly, networks trained with weight decay actually increase in accuracy when reducing the eosin stain intensity. This indicates that the difference between spectral decoupling and weight decay performance in Figure \ref{fig:real_world}D, may be partly due to differences in the stain intensities between the two medical centres. To explore this possibility, the stain intensities of the external dataset are normalized with the Macenko method \cite{macenko2009} to match the training data stain intensities and the resulted performance increases are reported in Figure \ref{fig:peso}C. Both networks trained with either spectral decoupling or weight decay benefit from stain normalization. Stain normalization is especially beneficial for networks trained with weight decay, where the mean network accuracy is increased by 7.5 percentage points. Networks trained with spectral decoupling still perform better than networks trained with weight decay coupled with stain normalization. These results demonstrate that spectral decoupling can complement or even replace normalization methods, with negligible computational requirements (Figure \ref{fig:peso}D).

\subsection{COVID-19 detection}
\label{sec:covid_detection}

To assess whether spectral decoupling can help in real-world situations with strong dominant features and spurious correlations, we train 5 networks with and without spectral decoupling to detect COVID-19 positive patients in chest radiographs. Two different training datasets are used to train the networks and all networks are evaluated on the same external validation set, described in Section \ref{dataset:covid}. We first train neural networks with the BIMCV$\pm$ dataset, which represents an ideal situation where both the positive and negative samples originate from similar sources. Second, we train networks with the combined PadChest and BIMCV$\pm$ dataset. This dataset represents a situation where the network can easily achieve high performance by only learning to detect where a sample originates as most of the negative samples come from a single medical centre.

\begin{figure}[t]
  \centering
  \includegraphics[width=\columnwidth]{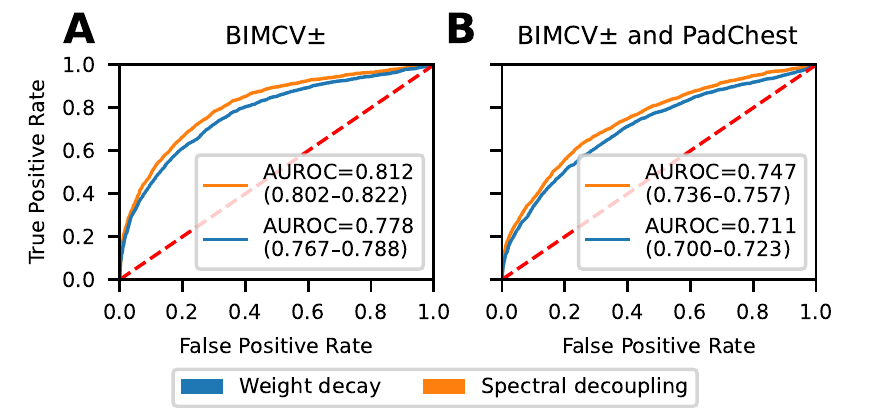}
  \caption{Receiver operating characteristic (ROC) curves for COVID-19 detection. Inset values indicate the areas under the ROC (AUROC) values and bootstrapped 95\% confidence intervals. Networks trained with spectral decoupling achieve significantly higher AUROC values compared to networks trained with weight decay.}
  \label{fig:covid}
\end{figure}

After training all networks, the predictions from each network are averaged to obtain ensemble predictions for both weight decay and spectral decoupling. ROC curves for ensemble predictions are presented in Figure \ref{fig:covid}, with bootstrapped ($n=1000$) 95\% confidence intervals (CI) for each area under the ROC curve (AUROC) value. Networks trained with spectral decoupling achieve significantly higher AUROC values for both BIMCV$\pm$ (DeLong's test: $Z=-15.914, p=10^{-56}$) and the combined PadChest and BIMCV$\pm$ (DeLong's test: $Z=-13.553, p=10^{-41}$) training datasets. On the BIMCV$\pm$ dataset, weight decay and spectral decoupling achieve AUROCs of 0.812 (95\% CI: 0.802 – 0.822) and 0.778 (95\% CI: 0.767 – 0.788), respectively. With the combined PadChest and BIMCV$\pm$ weight decay and spectral decoupling achieve AUROCs of 0.747 (95\% CI: 0.736 – 0.757) and 0.711 (95\% CI: 0.700 – 0.723), respectively.

When training networks with the combined PadChest and BIMCV$\pm$ dataset, AUROC values of networks trained with either method decrease, although the number of training samples is increased over tenfold. The decrease in AUROC is similar for both weight decay and spectral decoupling, 0.065 and 0.067, respectively. This indicates that spectral decoupling is unable to mitigate bias in the combined dataset. As most of the negative samples originate from a single medical centre, shortcut learning seems to happen even though spectral decoupling encourages the network to learn more features. Detecting where a sample originates is especially easy with radiographs due to systematic differences between data repositories and medical centres, which could be exploited by a neural network \cite{degrave2021covid}. Thus, the higher AUROC value of spectral decoupling is more likely due to increased robustness to data distribution shifts than avoidance of shortcut learning.

\section{Discussion}

Generalisation performance is defined as the main challenge standing in the way of true clinical adoption of a neural network \cite{van2021toclinic}. Van der Laak \emph{et al.}~\cite{van2021toclinic} argue that there is a need for public datasets which are truly representative of clinical practice. Although this is indeed important, we argue that training datasets, no matter how large, will never account for all possible variations caused by differences in imaging equipment, sample preparation and patient populations. Thus, it is crucial to couple extensive multi-source datasets with explicit and implicit bias mitigation methods to train neural networks which are robust to unseen variations.

Two explicit methods of bias mitigation have been proposed for medical imaging. Augmentation of the training samples is crucial as it substantially increases robustness for distribution shifts from the training data caused by differences in imaging equipment or sample preparation (Figure \ref{fig:robustness}, \cite{tellez2019quantifying}). Thus, it is strongly recommended to use extensive augmentation strategies for training neural networks intended for clinical practice. Normalization of all images to a common standard would substantially reduce the distribution shifts \cite{bel2019normalize, janowczyk2017stain, normalizing2020}, but comes with a considerable computational cost (Figure \ref{fig:peso}D). Both methods address important problems and should be complementary to any implicit methods of bias control.

Spectral decoupling is, to our knowledge, the first implicit bias mitigation method for addressing the generalisation issues in neural networks. The method is complementary to augmentation, increasing the robustness for distribution shifts already addressed with augmentation (Figure \ref{fig:robustness}A). Above all, spectral decoupling significantly increases the robustness for distribution shifts not addressed by augmentation (Figure \ref{fig:robustness}B) and could be used to replace computationally expensive stain normalisation methods (Figure \ref{fig:peso}C).

By encouraging the neural network to learn more features, spectral decoupling can also help in situations where the training dataset contains strong dominant features or spurious correlations (Table \ref{tab:sim_results}). This is crucial as the dominant features can also be inherent to the data, such as different cancer types. For example, with prostate cancer, different Gleason grades \cite{epstein2016gleason} are often unbalanced in the training set. Due to gradient starvation \cite{combes2018gs}, the features of the underrepresented Gleason grades may not be learned by the neural network. Balancing the dataset, so that all Gleason grades are represented equally, is not easy or even desired as the grading is based on a continuous range of histological patterns.

In COVID-19 detection, the networks' performance decreased similarly for both weight decay and spectral decoupling (Figure \ref{fig:covid}), when training the networks on the combined BIMCV$\pm$ and PadChest dataset. Radiographs contain systematic differences between data repositories and medical centres, such as laterality tokens and differences in the radiopacity of the image borders, which could arise from variations in patient position, radiographic projection or image processing \cite{degrave2021covid}. These differences can be easily leveraged by neural networks to detect where a single radiograph originates. We speculate that spectral decoupling was unable to prevent shortcut-learning due to the ease of shortcut learning in the combined PadChest and BIMCV$\pm$ dataset. In addition, our results showing the ability to prevent shortcut learning (Table \ref{tab:sim_results}) were obtained after considerable hyper-parameter optimization and no significant differences could be seen in the class activation maps between networks trained with either weight decay or spectral decoupling. Thus, removal of any obvious superficial correlations from the training dataset is crucial as there seems to be a limit of how much spectral decoupling can help with dominating features and spurious correlations.

The advantages of spectral decoupling can be clearly seen when the network is evaluated with out-of-distribution samples (Figures \ref{fig:robustness}, \ref{fig:real_world}) and \ref{fig:covid}). Neural networks trained with spectral decoupling retain their performance with samples further from the training data distribution, which is exactly what is required from neural networks intended for clinical practice \cite{van2021toclinic}. Although using an external dataset may not reveal all generalization problems, it is clear that without spectral decoupling the neural networks fail to generalize to this particular external dataset from Radboud University Medical Center (Figures \ref{fig:real_world}D and \ref{fig:peso}). Even in COVID-19 detection, where spectral decoupling seems to fail in preventing shortcut learning, the performance of the network is significantly increased over the state-of-the-art.


\section{Conclusions}

Spectral decoupling is the first implicit bias mitigation method for training neural networks to be used across multiple medical centres. The method adds no computational costs, is easy-to-implement, and complements and improves upon explicit bias mitigation methods. Our results recommend the use of spectral decoupling in all neural networks intended for clinical use.

\section*{Acknowledgments}

This work was supported by Cancer Foundation Finland [grant numbers 304667, 191118], Jane and Aatos Erkko Foundation [grant number 290520], Academy of Finland [grant number 322675] and Hospital District of Helsinki and Uusimaa [grant numbers TYH2018214, TYH2018222, TYH2019235, TYH2019249]. The authors also wish to acknowledge CSC – IT Center for Science, Finland, for generous computational resources and Wouter Bulten for directing us to the external prostate cancer dataset.

\section*{Ethics statement}

The images of the tissue slides are applied in this study based on national legislation and a research permission from the Helsinki University Hospital (§105).

\section*{Declaration of competing interests}

The authors have no interests to declare.

\section*{Credit authorship contribution statement}

\textbf{Joona Pohjonen:} Conceptualization; Data curation; Formal analysis; Investigation; Methodology; Project administration; Software; Validation; Visualization; Roles/Writing - original draft. \textbf{Carolin Stürenberg:} Data curation; Writing - review \& editing. \textbf{Antti Rannikko:} Funding acquisition; Resources; Supervision; Writing - review \& editing. \textbf{Tuomas Mirtti:} Data curation; Funding acquisition; Resources; Supervision; Writing - review \& editing. \textbf{Esa Pitkänen:} Supervision; Writing - review \& editing.

\bibliographystyle{unsrt}  
\bibliography{main}

\end{document}